\documentclass[preprint,prd,aps,showpacs,showkeys,nofootinbib]{revtex4}
\usepackage{graphicx}
\usepackage{dcolumn}
\usepackage{bm}
\usepackage{color}
\usepackage[dvipsnames]{xcolor}
\usepackage{amssymb}

\definecolor{black-blue}{RGB}{77,116,175}
\definecolor{black-yellow}{RGB}{231,162,33}
\definecolor{black-green}{RGB}{144,180,58}
\definecolor{black-red}{RGB}{246,95,50}

\begin{document}

\title{Lepton flavor violating decays $l_j\rightarrow l_i \gamma\gamma$}
\author{Ming-Yue Liu$^{1,2,3}$, Shu-Min Zhao$^{1,2,3}$\footnote{zhaosm@hbu.edu.cn}, Yi-Tong Wang$^{1,2,3}$, Xi Wang$^{1,2,3}$, Xin-Xin Long$^{1,2,3}$,\nonumber\\
 Tong-Tong Wang$^{1,2,3}$, Hai-Bin Zhang$^{1,2,3}$, Tai-Fu Feng$^{1,2,3,4}$.}

\affiliation{$^1$ Department of Physics, Hebei University, Baoding 071002, China}
\affiliation{$^2$ Hebei Key Laboratory of High-precision Computation and Application of Quantum Field Theory, Baoding, 071002, China}
\affiliation{$^3$ Hebei Research Center of the Basic Discpline for Computational Physics, Baoding, 071002, China}
\affiliation{$^4$ Department of Physics, Chongqing University, Chongqing 401331, China}
\date{\today}

\begin{abstract}
In this paper, we study the lepton flavor violating decays of the $l_j\rightarrow l_i \gamma\gamma$ (j=2, 3; i=1, 2) processes under the $U(1)_X$SSM. The $U(1)_X$SSM is the addition of three singlet new Higgs superfields and right-handed neutrinos to the minimal supersymmetric standard model (MSSM). Based on the latest experimental constraints of $l_j\rightarrow l_i \gamma\gamma$, we analyze the effects of different sensitive parameters on the results and made reasonable predictions for future experimental development. Numerical analysis shows that many parameters have a greater or lesser effect on lepton flavor violation(LFV), but the main sensitive parameters and sources leading to LFV are the non-diagonal elements involving the initial and final leptons. This work could provide a basis for the discovery of the existence of new physics (NP).

\end{abstract}

\keywords{lepton flavor violation, $U(1)_X$SSM, new physics.}

\maketitle

\section{Introduction}
The Standard Model (SM) contains many elementary particles, including fermions and bosons and their various forms. Although the SM has been relatively mature in previous developments, considering the lepton number in the SM, there is not LFV process in the SM \cite{p1}. However, the breaking theory of electric weak symmetry and neutrino oscillation experiment show that LFV exists both theoretically and experimentally \cite{neutrino1,neutrino2,neutrino3}, and the experimental observation of charged lepton flavor violation (cLFV) undoubtedly hints at the existence of new physics beyond neutrino oscillations \cite{cLFV}. Any sign of LFV in the experiment could be considered as evidence of LFV existence, and it is necessary to extend the SM through research. Afterwards, scientists extend the SM and obtain many extend models, among which the MSSM has been received much attention, but it is slowly discovered that there are problems in the MSSM such as the $\mu$ problem \cite{mu} and the zero mass neutrino \cite{neutrino4}. To break through these problems we notice the $U(1)$ extension of the MSSM, that is the extension of the MSSM with the $U(1)_X$ gauge group, and the symmetry group is $SU(3)_C\times SU(2)_L \times U(1)_Y\times U(1)_X$\cite{7,8,9}. It adds three singlet new Higgs superfields and right-handed neutrino superfields outside the MSSM \cite{10}, which relatively perfect solves the problem we face.

The latest upper limits on the LFV branching ratios of  $\mu\rightarrow e\gamma\gamma,~\tau\rightarrow\mu\gamma\gamma$  and  $\tau \rightarrow e\gamma\gamma $  at 90\% confidence level (CL) \cite{14} are:
\begin{eqnarray}
&&Br(\mu\rightarrow{e\gamma\gamma})<7.2\times10^{-11},\nonumber\\
&&Br(\tau\rightarrow{\mu\gamma\gamma})<5.8\times10^{-4},\nonumber\\
&&Br(\tau\rightarrow{e\gamma\gamma})<2.5\times10^{-4}.
\end{eqnarray}

In previous work\cite{11,2007,12}, studies have been carried out including a valid field theory analysis to correlate the charged lepton flavor violating processes $l_i\rightarrow l_j \gamma\gamma$ and $l_i\rightarrow l_j \gamma$. Model-independent upper bounds on the $l_i\rightarrow l_j \gamma\gamma$ rates are derived using the current upper bounds on the $l_i\rightarrow l_j \gamma$ rates \cite{11} and in the framework of effective field theory. The magnitude of the branching ratios of $\tau \rightarrow \mu\gamma$ and $\tau \rightarrow \mu\gamma\gamma$ decays caused by a lepton flavor violating Higgs interaction $H\tau$$\mu$ is studied \cite{2007}. Here we will conduct a more comprehensive study of the LFV of the $l_j\rightarrow l_i \gamma\gamma$ under the $U(1)_X$SSM. We learn about the LFV of the $l_j\rightarrow l_i \gamma$ process \cite{12}, and its numerical results show that the experimental limit of the $l_j\rightarrow l_i \gamma$  is the most stringent for the parameter space constraint of the $U(1)_X$SSM. We study the LFV process of the $l_j\rightarrow l_i \gamma\gamma$  under the $U(1)_X$SSM in depth on this basis. $l_j\rightarrow l_i \gamma\gamma$ is more complicated and more difficult to study than $l_j\rightarrow l_i \gamma$. Compared with $l_j\rightarrow l_i \gamma$, the Feynman diagrams of $l_j\rightarrow l_i \gamma\gamma$ become more numerous and each of them becomes more complex.
We compare the LFVs of the $l_j\rightarrow l_i \gamma$ and $l_j\rightarrow l_i \gamma\gamma$ processes in the numerical section to make our study more interesting and to visualize the correlation between the two type processes.
The $l_j\rightarrow l_i \gamma\gamma$ process is a challenging but interesting process. We have derived and numerically analyzed the relevant Feynman diagrams. From the numerical results, we obtain reasonable parameter spaces. The effects of different reasonable parameters on the branching ratio Br($l_j\rightarrow l_i \gamma\gamma$) are compared.

This paper will expand according to the following structure. In Sec.II, we briefly introduce the essential content of the $U (1)_X$SSM, including its superpotential, the general soft breaking terms, the rotations and interactions of the eigenstates 'EWSB'. In Sec.III, we provide analytical expressions for the branching ratio of the $l_j \rightarrow l_i \gamma\gamma$ decay in the $U(1)_X$SSM. In Sec.IV, we give the corresponding parameters and numerical analysis. In Sec.V, we present a summary of this article.

\section{The essential  content of  $U(1)_X$SSM}
$U(1)_X$SSM is a $U(1)$ extension on the basis of MSSM, whose local gauge group is $SU(3)_C\times SU(2)_L \times U(1)_Y\times U(1)_X$ \cite{16,17,18,19}. The $U(1)_X$SSM mainly consists of the superpotential, rotations and interactions for eigenstates 'EWSB'
etc. Compared to MSSM, there are new superfields in $U(1)_X$SSM, such as right-handed neutrinos $\hat{\nu}_i$ and three Higgs singlets $\hat{\eta},~\hat{\bar{\eta}},~\hat{S}$. The representation of the superpotential in the $U(1)_X$SSM is:
\begin{eqnarray}
&&W=l_W\hat{S}+\mu\hat{H}_u\hat{H}_d+M_S\hat{S}\hat{S}-Y_d\hat{d}\hat{q}\hat{H}_d-Y_e\hat{e}\hat{l}\hat{H}_d+\lambda_H\hat{S}\hat{H}_u\hat{H}_d
\nonumber\\&&\hspace{0.6cm}+\lambda_C\hat{S}\hat{\eta}\hat{\bar{\eta}}+\frac{\kappa}{3}\hat{S}\hat{S}\hat{S}+Y_u\hat{u}\hat{q}\hat{H}_u+Y_X\hat{\nu}\hat{\bar{\eta}}\hat{\nu}
+Y_\nu\hat{\nu}\hat{l}\hat{H}_u.
\end{eqnarray}

In the above equation, the vacuum expectation values(VEVs) of the two Higgs doublet states $H_{u},H_{d}$ are $v_u,~v_d$ and the VEVs of the three Higgs singlet states $\eta$, $\bar{\eta}$, S are $~v_\eta$, $v_{\bar\eta}$ and $v_S$ respectively. The Higgs superfields are displayed as follows:
\begin{eqnarray}
&&\hspace{1cm}H_{u}=\left(\begin{array}{c}H_{u}^+\\{1\over\sqrt{2}}\Big(v_{u}+H_{u}^0+iP_{u}^0\Big)\end{array}\right),
~~~~~~
H_{d}=\left(\begin{array}{c}{1\over\sqrt{2}}\Big(v_{d}+H_{d}^0+iP_{d}^0\Big)\\H_{d}^-\end{array}\right),
\nonumber\\&&\eta={1\over\sqrt{2}}\Big(v_{\eta}+\phi_{\eta}^0+iP_{\eta}^0\Big),~~~
\bar{\eta}={1\over\sqrt{2}}\Big(v_{\bar{\eta}}+\phi_{\bar{\eta}}^0+iP_{\bar{\eta}}^0\Big),~~
S={1\over\sqrt{2}}\Big(v_{S}+\phi_{S}^0+iP_{S}^0\Big).
\end{eqnarray}
There are two angles $\tan\beta$ and $\tan\beta_\eta$, which are defined as  $\tan\beta=v_u/v_d$  and  $\tan\beta_\eta=v_{\bar{\eta}}/v_{\eta}$.  The soft SUSY breaking terms of $U(1)_X$SSM are shown as:
\begin{eqnarray}
&&\mathcal{L}_{soft}=\mathcal{L}_{soft}^{MSSM}-B_SS^2-L_SS-\frac{T_\kappa}{3}S^3-T_{\lambda_C}S\eta\bar{\eta}
+\epsilon_{ij}T_{\lambda_H}SH_d^iH_u^j\nonumber\\&&\hspace{1cm}
-T_X^{IJ}\bar{\eta}\tilde{\nu}_R^{*I}\tilde{\nu}_R^{*J}
+\epsilon_{ij}T^{IJ}_{\nu}H_u^i\tilde{\nu}_R^{I*}\tilde{l}_j^J
-m_{\eta}^2|\eta|^2-m_{\bar{\eta}}^2|\bar{\eta}|^2-m_S^2S^2\nonumber\\&&\hspace{1cm}
-(m_{\tilde{\nu}_R}^2)^{IJ}\tilde{\nu}_R^{I*}\tilde{\nu}_R^{J}
-\frac{1}{2}\Big(M_S\lambda^2_{\tilde{X}}+2M_{BB^\prime}\lambda_{\tilde{B}}\lambda_{\tilde{X}}\Big)+h.c~.
\end{eqnarray}

\begin{table}
\caption{ The superfields in $U(1)_X$SSM}
\begin{tabular}{|c|c|c|c|c|c|c|c|c|c|c|c|}
\hline
Superfields & $\hspace{0.1cm}\hat{q}_i\hspace{0.1cm}$ & $\hat{u}^c_i$ & $\hspace{0.2cm}\hat{d}^c_i\hspace{0.2cm}$ & $\hat{l}_i$ & $\hspace{0.2cm}\hat{e}^c_i\hspace{0.2cm}$ & $\hat{\nu}_i$ & $\hspace{0.1cm}\hat{H}_u\hspace{0.1cm}$ & $\hat{H}_d$ & $\hspace{0.2cm}\hat{\eta}\hspace{0.2cm}$ & $\hspace{0.2cm}\hat{\bar{\eta}}\hspace{0.2cm}$ & $\hspace{0.2cm}\hat{S}\hspace{0.2cm}$ \\
\hline
$SU(3)_C$ & 3 & $\bar{3}$ & $\bar{3}$ & 1 & 1 & 1 & 1 & 1 & 1 & 1 & 1  \\
\hline
$SU(2)_L$ & 2 & 1 & 1 & 2 & 1 & 1 & 2 & 2 & 1 & 1 & 1  \\
\hline
$U(1)_Y$ & 1/6 & -2/3 & 1/3 & -1/2 & 1 & 0 & 1/2 & -1/2 & 0 & 0 & 0  \\
\hline
$U(1)_X$ & 0 & -1/2 & 1/2 & 0 & 1/2 & -1/2 & 1/2 & -1/2 & -1 & 1 & 0  \\
\hline
\end{tabular}
\label{JJ1}
\end{table}

The particle content and charge assignments for $U(1)_X$SSM are shown in the Table \ref {JJ1}. The new effect of  the gauge kinetic mixing in $U(1)_X$SSM has never been seen before in MSSM. Here $U(1)_Y$ and $U(1)_X$ are two Abelian groups, and we denote $U(1)_Y$ charge by $Y^Y$ and $U(1)_X$ charge by $Y^X$. They generate the  gauge kinetic mixing. The rotations of eigenstates 'EWSB' are divided into two categories. One is rotation in mass sector and the other is rotation in gauge sector. Rotations in gauge sector for eigenstates 'EWSB' are:
\begin{eqnarray}
&&\left(\begin{array}{ccc}B_{\rho}\\W_{3\rho}\\V_{B_X}\\\end{array}\right)={Z^{\gamma ZZ{'}}}\left(\begin{array}{ccc}\gamma_{\rho}\\Z_{\rho}\\Z{'}_{\rho}\\\end{array}\right),
~~~\left(\begin{array}{ccc}\lambda_{\tilde{W},1}\\\lambda_{\tilde{W},2}\\\lambda_{\tilde{W},3}\\\end{array}\right)=
{Z^{\tilde{W}}}\left(\begin{array}{ccc}\tilde{W}^{-}\\\tilde{W}^{+}\\\tilde{W}^{0}\\\end{array}\right),
\nonumber\\&&\hspace{3cm}
\left(\begin{array}{ccc}W_{1\rho}\\W_{2\rho}\\\end{array}\right)={Z^{W}}\left(\begin{array}{ccc}W^{-}_{\rho}\\W^{-}_{\rho}\\\end{array}\right).
\end{eqnarray}
$\theta_{W}$ is the Weinberg angle, the mixing matrices are parametrized by:
\begin{eqnarray}
&&Z^{\gamma ZZ{'}}=\left(\begin{array}{ccc}
\cos\theta_{W}&-\cos\theta_{W}' \sin\theta_{W}&\sin\theta_{W} \sin\theta_{W}'\\
\sin\theta_{W}&\cos\theta_{W} \sin\theta_{W}'&-\cos\theta_{W} \sin\theta_{W}'\\
0&\sin\theta_{W}'&\cos\theta_{W}'\\
\end{array}\right),\nonumber\\&&\hspace{0.5cm}
Z^W=\left(\begin{array}{ccc}
\frac{1}{\sqrt{2}}&\frac{1}{\sqrt{2}}\\
-i \frac{1}{\sqrt{2}}&i \frac{1}{\sqrt{2}}\\
\end{array}\right),
~~Z^{\tilde{W}}=\left(\begin{array}{ccc}
\frac{1}{\sqrt{2}}&\frac{1}{\sqrt{2}}&0\\
-i \frac{1}{\sqrt{2}}&i \frac{1}{\sqrt{2}}&0\\
0&0&1\\
\end{array}\right).
\end{eqnarray}

There are mass matrices for scalars and fermions. The mass squared matrix for $CP$-odd sneutrino $(\sigma_{l}, \sigma_{r})$ reads:
\begin{eqnarray}
M^2_{\tilde{\nu}^I} = \left(
\begin{array}{cc}
m_{{\sigma}_{l}{\sigma}_{l}} &m^T_{{\sigma}_{r}{\sigma}_{l}}\\
m_{{\sigma}_{l}{\sigma}_{r}} &m_{{\sigma}_{r}{\sigma}_{r}}\end{array}
\right),
 \end{eqnarray}
\begin{eqnarray}
&&m_{{\sigma}_{l}{\sigma}_{l}}= \frac{1}{8} \Big((g_{1}^{2} + g_{Y X}^{2} + g_{2}^{2}+  g_{Y X} g_{X})( v_{d}^{2}- v_{u}^{2})
+  2g_{Y X} g_{X}(v_{\eta}^{2}-v_{\bar{\eta}}^{2})\Big)
\nonumber\\&&\hspace{1.8cm}+\frac{1}{2} v_{u}^{2}{Y_{\nu}^{T}  Y_\nu}  + M_{\tilde{L}}^2,
 \\&&m_{{\sigma}_{l}{\sigma}_{r}} = \frac{1}{\sqrt{2} } v_uT_\nu -  v_u v_{\bar{\eta}} {Y_X  Y_\nu}
  - \frac{1}{2}v_d ({\lambda}_{H}v_S  + \sqrt{2} \mu )Y_\nu,\\&&
m_{{\sigma}_{r}{\sigma}_{r}}= \frac{1}{8} \Big((g_{Y X} g_{X}+g_{X}^{2})(v_{d}^{2}- v_{u}^{2})
+2g_{X}^{2}(v_{\eta}^{2}- v_{\bar{\eta}}^{2})\Big)- v_{\eta} v_S Y_X {\lambda}_{C}\nonumber \\&&\hspace{1.8cm}
+M_{\tilde{\nu}}^2 + \frac{1}{2} v_{u}^{2}|Y_\nu|^2+  v_{\bar{\eta}} (2 v_{\bar{\eta}}Y_X  Y_X  - \sqrt{2} T_X).
\end{eqnarray}
The mass squared matrix for $CP$-even sneutrino $({\phi}_{l}, {\phi}_{r})$ reads:
\begin{eqnarray}
M^2_{\tilde{\nu}^R} = \left(
\begin{array}{cc}
m_{{\phi}_{l}{\phi}_{l}} &m^T_{{\phi}_{r}{\phi}_{l}}\\
m_{{\phi}_{l}{\phi}_{r}} &m_{{\phi}_{r}{\phi}_{r}}\end{array}
\right),\label{Rsneu}
 \end{eqnarray}
\begin{eqnarray}
&&m_{{\phi}_{l}{\phi}_{l}}= \frac{1}{8} \Big((g_{1}^{2} + g_{Y X}^{2} + g_{2}^{2}+ g_{Y X} g_{X})( v_{d}^{2}- v_{u}^{2})
+  g_{Y X} g_{X}(2 v_{\eta}^{2}-2 v_{\bar{\eta}}^{2})\Big)
\nonumber\\&&\hspace{1.8cm}+\frac{1}{2} v_{u}^{2}{Y_{\nu}^{T}  Y_\nu}  + M_{\tilde{L}}^2,
 \\&&m_{{\phi}_{l}{\phi}_{r}} = \frac{1}{\sqrt{2} } v_uT_\nu  +  v_u v_{\bar{\eta}} {Y_X  Y_\nu}
  - \frac{1}{2}v_d ({\lambda}_{H}v_S  + \sqrt{2} \mu )Y_\nu,\\&&
m_{{\phi}_{r}{\phi}_{r}}= \frac{1}{8} \Big((g_{Y X} g_{X}+g_{X}^{2})(v_{d}^{2}- v_{u}^{2})
+2g_{X}^{2}(v_{\eta}^{2}- v_{\bar{\eta}}^{2})\Big) + v_{\eta} v_S Y_X {\lambda}_{C}\nonumber \\&&\hspace{1.8cm}
 +M_{\tilde{\nu}}^2 + \frac{1}{2} v_{u}^{2}|Y_\nu|^2+  v_{\bar{\eta}} (2 v_{\bar{\eta}}Y_X  Y_X  + \sqrt{2} T_X).
\end{eqnarray}
The mass matrix for chargino reads:
\begin{eqnarray}
m_{\tilde{\chi}^-} = \left(
\begin{array}{cc}
M_2&\frac{1}{\sqrt{2}}g_2v_\mu\\
\frac{1}{\sqrt{2}}g_2v_d&\frac{1}{\sqrt{2}}\lambda_{H} v_S+\mu\end{array}
\right).
 \end{eqnarray}
This matrix is diagonalized by U and V:\begin{eqnarray}
U^{*} m_{\tilde{\chi}^-} V^{\dagger}= m_{\tilde{\chi}^-}^{diag},
 \end{eqnarray}
with
\begin{eqnarray}
&&{\tilde{W}^-}=\sum_{t_2}U_{j1}^{*} \lambda_{j}^-,~~~~~~{\tilde{H}_{d}^-}=\sum_{t_2}U_{j2}^{*} \lambda_{j}^-,\nonumber\\
&&{\tilde{W}^+}=\sum_{t_2}V_{1j}^{*} \lambda_{j}^+,~~~~~~{\tilde{H}_{u}^-}=\sum_{t_2}V_{2j}^{*} \lambda_{j}^+.
 \end{eqnarray}

Furthermore, some other required mass matrices can be found in Refs.\cite{16,17}. There are interactions between the eigenstates 'EWSB', and here we list some of the couplings required in the $U(1)_X$SSM. We derive the vertices of lepton-chargino-sneutrino:
\begin{eqnarray}
&&\mathcal{L}_{\bar{l}\chi^-\tilde{\nu}^I}=\frac{i}{\sqrt{2}}\bar{l}_i\Big\{U_{j2}^*Z_{ki}^{I*}Y_l^iP_L-g_2V_{j1}Z_{ki}^{I*}P_R\Big\}\chi_j^-\tilde{\nu}_k^I,
\nonumber\\&&\hspace{0cm}\mathcal{L}_{\bar{l}\chi^-\tilde{\nu}^R}=\frac{1}{\sqrt{2}}\bar{l}_i\Big\{U_{j2}^*Z_{ki}^{R*}Y_l^iP_L-g_2V_{j1}Z_{ki}^{R*}P_R\Big\}\chi_j^-\tilde{\nu}_k^R.\label{S1}
\end{eqnarray}
We also derive the vertex of  neutralino-lepton-slepton:
\begin{eqnarray}
&&\mathcal{L}_{\bar{\chi}^0l\tilde{L}}=\bar{\chi}_i^0\Big\{\Big(\frac{1}{\sqrt{2}}(g_1N_{i1}^*+g_2N_{i2}^*+g_{YX}N_{i5}^*)Z_{kj}^E-N_{i3}^*Y_l^jZ_{K(3+j)}^E\Big)P_L,
\nonumber\\&&\hspace{1.5cm}-\Big[\frac{1}{\sqrt{2}}\Big(2g_1N_{i1}+(2g_{YX}+g_X)N_{i5}\Big)Z_{k(3+a)}^E+Y_l^jZ_{kj}^EN_{i3}\Big]P_R\Big\}l_j\tilde{L}_k.\label{S2}
\end{eqnarray}

There are some other vertices being needed, and to save space in this text, the remaining vertices can be found in Refs.\cite{18,19,20,21}.

\section {formulation}
\begin{figure}[ht]
\setlength{\unitlength}{5.0mm}
\centering
\includegraphics[width=6.5in]{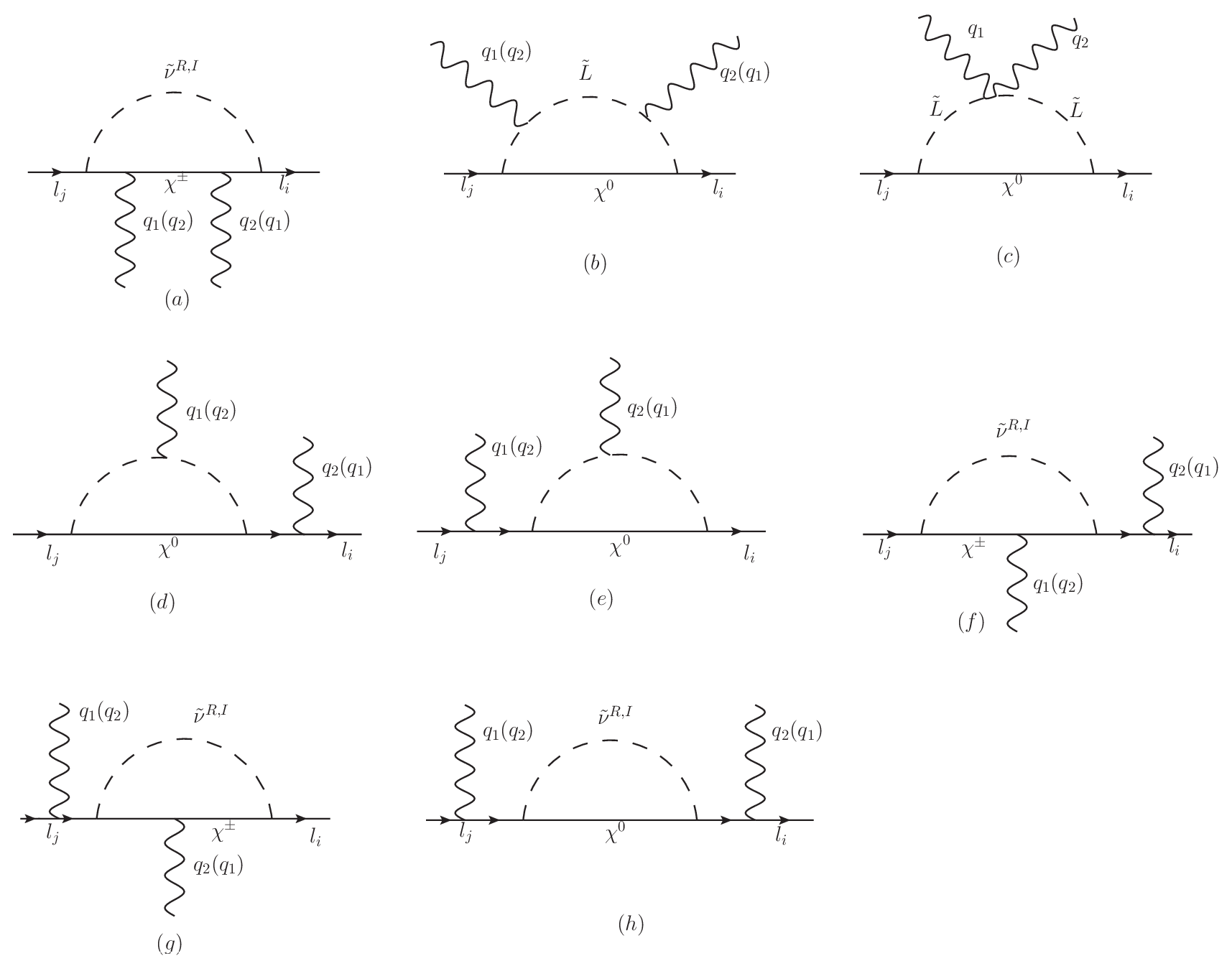}
\caption{  Feynman diagrams for the $l_j\rightarrow{l_i\gamma\gamma}$ processes in the $U(1)_X$SSM.}\label{N1}
\end{figure}

In this section, we study the amplitude and branching ratio of $l_j\rightarrow l_i \gamma\gamma$ (j = 2,3; i = 1,2).
Here, we do not extract the operators as in the previous work \cite{12,14},
but do a more comprehensive process. The corresponding Feynman diagrams are shown
in Fig.\ref {N1}. We study two-photon with $q_1$ and $q_2$ representing
the momentums of the two photons, respectively. In  Fig.\ref{N1} all diagrams
except Fig.\ref {N1}(c) contain diagrams with $q_1$ on the left and $q_2$ on
 the right and with the positions of $q_1$ and $q_2$ switched.

We choose a Feynman diagram in Fig.\ref{N1} to analyze. The Feynman amplitude of Fig.\ref{N1}(a) is:
\begin{eqnarray}
&&\mathcal{M}_{(a)}=\bar{U}_i(p+q_1+q_2)\int\frac{d^Dk}{(2\pi)^D}\frac{1}{[(k+p+q_1+q_2)^2-m_f^2][(k+p+q_1)^2-m_f^2][(p+k)^2-m_f^2]}
\nonumber\\&&\hspace{1.5cm}\frac{1}{(k^2-m_s^2)} \Big((A_LP_L+A_RP_R)({k\!\!\!\slash+p\!\!\!\slash+{q_1}\!\!\!\!\!\slash+{q_2}\!\!\!\!\!\slash+m_f})\gamma^\nu e ({k\!\!\!\slash+p\!\!\!\slash+{q_1}\!\!\!\!\!\slash+m_f})\gamma^\mu e
\nonumber\\&&\hspace{1.5cm}({k\!\!\!\slash+p\!\!\!\slash+m_f})(B_LP_L+B_RP_R)\Big)U_j(p)\varepsilon_\mu^*(q_1)\varepsilon_\nu^*(q_2),
\end{eqnarray}
where p is the injected lepton momentum, $m_f$ corresponds to the chargino mass, $m_s$ corresponds to the scalar neutrino ($CP$-even or $CP$-odd) mass. $A_L, B_L, A_R$ and $B_R$ represent the coupling vertices we mentioned in Sec.II. $\mathcal{L}=\bar{F_1}(iA_L+iA_R)F_2S$, $\bar{U}_i(p+q_1+q_2)$ and $U_j(p)$ are the wave functions of the external leptons.

Since the mass of the inner particle is large and the mass of the outer particle is small. So, for the denominator term in $\mathcal{M}_{(a)}$ we can do the following expansion to simplify the calculation
\begin{eqnarray}
&&\frac{1}{(k+p)^2-m_f^2}=\frac{1}{k^2-m_f^2}(1-\frac{p^2+2k\cdot p}{k^2-m_f^2}+\frac{4(k\cdot p)^2}{(k^2-m_f^2)^2}).
\end{eqnarray}

We take all the diagrams in Fig.\ref {N1} as above and further simplify them. Since the results produced by its computational process are very large, we use
$Mathematica \ll HighEnergyPhysics`FeynCalc`$ for analytical calculations.
We derive the Feynman amplitudes of all the diagrams, using the on-shell condition, so that $q_1^2=0$, $q_2^2=0$, $p^2=m_j^2$. Summation is performed and its mode square is calculated as $|\mathcal{M}|^2$. In $|\mathcal{M}|^2$ there are various loop functions such as $A_LB_R\frac{k^2}{(k^2-m_s^2)^3(k^2-m_f^2)}m_f^2$, $A_RB_L\frac{k^4}{(k^2-m_s^2)^4(k^2-m_f^2)}m_f^3$, $A_RB_L\frac{1}{(k^2-m_s^2)(k^2-m_f^2)^3}m_f^2$ and so on, make $x$=$\frac{m_s^2}{m_\omega^2}$, $y$=$\frac{m_f^2}{m_\omega^2}$.

Here:
\begin{eqnarray}
&&\int\frac{d^Dk}{(2\pi)^D}\frac{k^2}{(k^2-m_s^2)^3(k^2-m_f^2)}=\frac{i}{32 \pi ^2}{\left(\frac{2 y^2 (\log (x)+\log (y))}{(y-x)^3}+\frac{3 y-x}{(x-y)^2}\right)},\nonumber\\
&&\int\frac{d^Dk}{(2\pi)^D}\frac{k^4}{(k^2-m_s^2)^4(k^2-m_f^2)}=\frac{i}{96\pi ^2}{\left(\frac{7 x y-2 x^2-11 y^2}{(x-y)^3}+\frac{6 y^3 (\log (x)-\log (y))}{(x-y)^4}\right)},\nonumber\\
&&\int\frac{d^Dk}{(2\pi)^D}\frac{1}{(k^2-m_s^2)(k^2-m_f^2)^3}=\frac{i}{32 \pi ^2}{\left(\frac{x+y}{y (x-y)^2}-\frac{2 x (\log (x)-\log (y))}{(x-y)^3}\right)}.
\end{eqnarray}

\begin{figure}[ht]
\setlength{\unitlength}{2.5mm}
\centering
\includegraphics[width=4in]{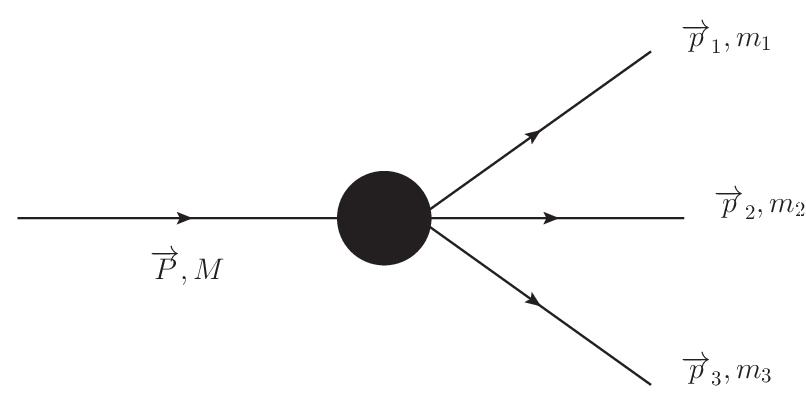}
\caption{Definitions of variables for three-body decays.}\label{N2}
\end{figure}

In order to obtain the branching ratio, we derive the three-body decay (Fig.\ref {N2}) by defining  $p_i+q_j=p_{ij}$, $p_{ij}^2=m_{ij}^2$. With total energy $E$, the momentums of the three final state particles are in the same plane and their relative directions are fixed. Let their Euler angles be $(\alpha, \beta, \gamma)$ to determine the final system with respect to the initial orientation of the particles, we obtain:
\begin{eqnarray}
&&d\Gamma=\frac{1}{(2\pi)^5}\frac{1}{16M}|\mathcal{M}|^2dE_1dE_3d \alpha d(\cos\beta)d\gamma.
\end{eqnarray}
If the decaying particle is a scalar or we average over its spin state, then the integration in terms of the equation is:
\begin{eqnarray}
&&d\Gamma=\frac{1}{(2\pi)^3}\frac{1}{8M}|\mathcal{M}|^2dE_1dE_3.
\end{eqnarray}

We integrate the $E_1$, $E_2$ in $|\mathcal{M}|^2$, $0\leq E_1\leq \frac{m_j}{2}$, $\frac{m_j}{2}-E_1\leq E_2\leq\frac{m_j}{2} $. We obtain the decay width and branching ratio as:
\begin{eqnarray}
&&\Gamma(l_j\rightarrow l_i\gamma\gamma)=\frac{1}{(2\pi)^3}\frac{1}{8M}\int|\mathcal{M}|^2dE_1dE_2,
\end{eqnarray}
\begin{eqnarray}
&&Br(l_j\rightarrow l_i\gamma\gamma)=\frac{\Gamma(l_j\rightarrow l_i\gamma\gamma)}{\Gamma_{l_j}}.
\end{eqnarray}

\section{Numerical analysis}
In this section, we perform numerical results considering experimental constraints on the lepton flavor violating process $l_j\rightarrow{l_i\gamma\gamma}$ and the lightest $CP$-even Higgs mass $m_{h0}=125.25$ GeV \cite{14,22,23}. We need to find some sensitive parameters from the parameters used, so as to obtain reasonable numerical results. Since the experimental limit of the $l_j\rightarrow{l_i\gamma}$ \cite{12} process has the most stringent parameter space constraint on the $U(1)_X$SSM, we need to consider the effect of $l_j\rightarrow{l_i\gamma}$  on the LFV. The strongest limit of $\mu\rightarrow e\gamma$, and the other limits can be satisfied if the limit of $\mu\rightarrow e\gamma$ is satisfied \cite{16}. So we plot $l_j\rightarrow{l_i\gamma}$ processes with the same parameters as the $l_j\rightarrow{l_i\gamma\gamma}$ processes, so that we can ensure that the studied processes strictly satisfy the $l_j\rightarrow{l_i\gamma}$ constraints.
According to the latest data from the LHC , we take for the scalar lepton mass greater than 700 GeV, the chargino mass greater than 1100 GeV, for more detailed limits refer to Refs.\cite{24,25,26,27,28,29}. We will discuss the processes $\mu\rightarrow e\gamma\gamma$, $\tau\rightarrow \mu\gamma\gamma$, $\tau\rightarrow e\gamma\gamma$ in three subsections and plot the relational and scatter diagrams with different parameters. By analyzing these plots and the experimental limits of the branching ratios, a reasonable parameter space is found to explain the LFV.

In summary, considering the experimental constraints described above, we adopt the following parameters in the numerical calculation.
\begin{eqnarray}
&&\mu=M_{BL}=T_{\lambda_C}=T_{\lambda_H}=T_\kappa=1~{\rm TeV},~~M_{BB'}=M_S=0.4~{\rm TeV},~~\lambda_H=0.1,\nonumber\\
&&l_W=B_\mu=B_S=0.1~{\rm TeV}^2,~~T_{Xii}=-1~{\rm TeV},~~\kappa=0.1,~~Y_{Xii}=1~{\rm TeV}(i=1,2,3),\nonumber\\
&&M_{\tilde{E}ii}^2=0.8~{\rm TeV}^2,~~M_{\tilde{\nu}ii}^2=0.3~{\rm TeV}^2,~~T_{\tilde{e}ii}=0.5~{\rm TeV},~\lambda_C=-0.25.
\end{eqnarray}

To simplify the numerical study, we use the relation of the parameters, which vary in the numerical analysis below
\begin{eqnarray}
&&M_{\tilde{L}ij}^2=M_{\tilde{L}ji}^2,~~T_{eij}=T_{eji},~~M_{\tilde{E}ii}^2=M_{\tilde{E}}^2,~~M_{\tilde{L}ii}^2=M_{\tilde{L}}^2,~~M_{\tilde{\nu}ij}^2=M_{\tilde{\nu}ji}^2,\nonumber\\
&&M_{\tilde{E}ij}^2=M_{\tilde{E}ji}^2,~~g_{YX},~~M_2,~~\tan\beta,~~g_X,~(i,j=1,2,3,~i\neq j).
\end{eqnarray}

Normally, the non-diagonal elements of the parameters are defined to be zero, unless we specify otherwise.
\subsection{$\mu\rightarrow e\gamma\gamma$}
\begin{figure}[ht]
\setlength{\unitlength}{5mm}
\centering
\includegraphics[width=2.9in]{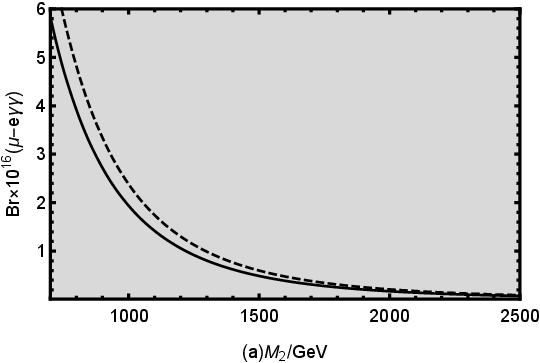}
\setlength{\unitlength}{5mm}
\centering
\includegraphics[width=2.9in]{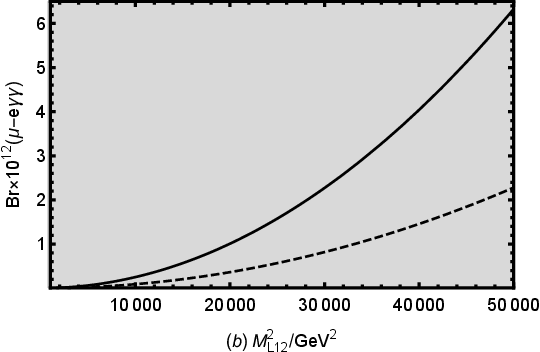}
\setlength{\unitlength}{5mm}
\centering\nonumber\\
\includegraphics[width=2.8in]{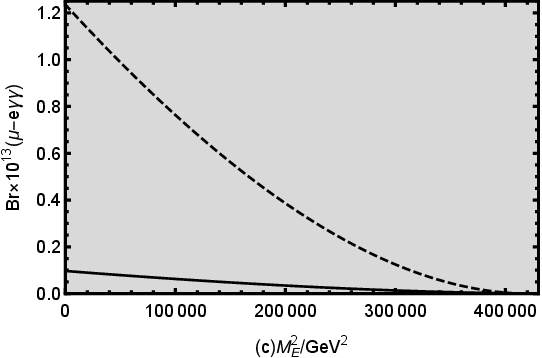}
\setlength{\unitlength}{5mm}
\centering
\includegraphics[width=2.9in]{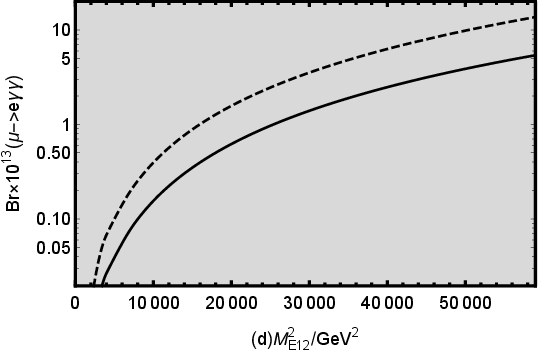}
\setlength{\unitlength}{5mm}
\centering\nonumber\\
\includegraphics[width=2.9in]{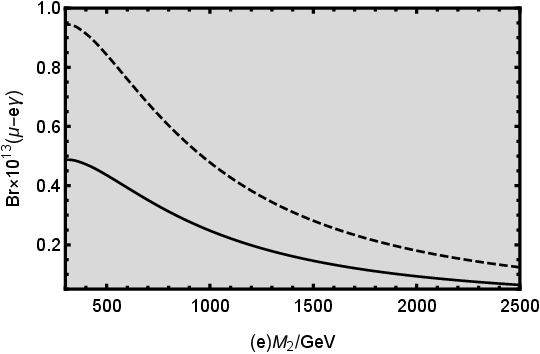}
\setlength{\unitlength}{5mm}
\centering
\includegraphics[width=2.9in]{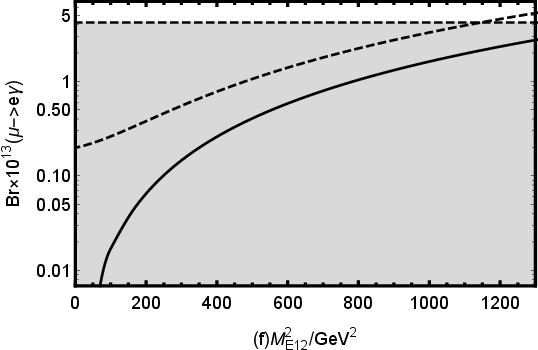}
\caption{Br($\mu\rightarrow e\gamma\gamma$) and Br($\mu\rightarrow e\gamma$) schematic diagrams affected by different parameters. The gray areas are reasonable value range, where Br($\mu\rightarrow e\gamma\gamma$) and Br($\mu\rightarrow e\gamma$) are lower than the upper limit. The dashed and solid lines in Fig.\ref{T1}(a)(e) correspond to $\tan\beta=25$ and $\tan\beta=20$. The dashed and solid lines in Fig.\ref{T1}(b) correspond to $M_2=1600~{\rm GeV}$ and $M_2=1200~{\rm GeV}$. The dashed and solid lines in Fig.\ref{T1}(c) correspond to $T_{e13}=200~{\rm GeV}$ and $T_{e13}=100~{\rm GeV}$. In Fig.\ref{T1}(d)(f), the $g_{YX}=0.1$ (dashed line) and $g_{YX}=0.2$ (solid line).}{\label {T1}}
\end{figure}

We perform numerical calculation for Br($\mu\rightarrow e\gamma\gamma$) and plot the relationship and scatter diagrams for different parameters in order to clearly show the numerical results. In Fig.\ref{T1} the gray area is the experimental limit satisfied by the processes.

With the parameters $M_{\tilde{L}}^2=1\times10^6~{\rm GeV}^2$, $g_X=0.3$, we plot Br($\mu\rightarrow e\gamma\gamma$) and Br($\mu\rightarrow e\gamma$) versus $M_2$ in the Fig.\ref{T1}(a)(e). The dashed lines correspond to $\tan\beta$=25 and the solid lines correspond to $\tan\beta$=20. We find that lines decrease with increasing $M_2$ in the range of 700~{\rm GeV}-2500~{\rm GeV}. The dashed lines are larger than the solid lines, and both the solid and dashed lines are located in the gray area indicating that both are within the experimental limits. In the Fig.\ref{T1}(b), we plot Br($\mu\rightarrow e\gamma\gamma$) versus $M_{\tilde{L}12}^2$, in which the dashed line corresponds to $M_2$=1600~{\rm GeV} and the solid line corresponds to $M_2$=1200~{\rm GeV}. We find that two lines increase with increasing $M_{\tilde{L}12}^2$ in the range of $0- 5\times10^4~{\rm GeV}^2$. The solid line is larger than the dashed line, and both the solid and dashed lines are located in the gray area satisfying the experimental limits.

In the Fig.\ref{T1}(c), we plot Br($\mu\rightarrow e\gamma\gamma$) versus $M_{\tilde{E}}^2$, in which the dashed line corresponds to $T_{e13}$=200~{\rm GeV} and the solid line corresponds to $T_{e13}$=100~{\rm GeV}. We find that two lines decrease with increasing $M_{\tilde{E}}^2$ in the range of $0-4\times10^5~{\rm GeV}^2$. The dashed line is larger than the solid line. Both the solid and dashed lines are located in the gray area indicating that both are within the experimental limits. In Fig.\ref{T1}(d)(f) we let $\tan\beta=20$, $M_2=1200$~{\rm GeV}, $M_E^2=8\times10^5~{\rm GeV^2}$, and plot Br($\mu\rightarrow e\gamma\gamma$) and Br($\mu\rightarrow e\gamma$) versus $M_{E12}^2$, with $g_{YX}=0.1$ (dashed line), $g_{YX}=0.2$ (solid line). It is clear that, both lines increase gradually, meaning that Br($\mu\rightarrow e\gamma\gamma$) and Br($\mu\rightarrow e\gamma$) get larger as $M_{E12}^2$ increase. As solid and dashed lines go from bottom to top,  Br($\mu\rightarrow e\gamma\gamma$) and Br($\mu\rightarrow e\gamma$) increases as $g_{YX}$ decreases.

$M_{\tilde{L}12}^2$ and $M_{\tilde{E}}^2$ are flavor parameters that appear in the slepton, $CP$-even sneutrino and $CP$-odd sneutrino mass matrices. Br($\mu\rightarrow e\gamma\gamma$) decreases with increasing parameters $M_2$ and $M_{\tilde{E}}^2$ and increases with increasing $M_{\tilde{L}12}^2$. In Fig.\ref{T1}(a)(c), the slopes of the dashed lines are higher than the solid lines, meaning that the larger $\tan\beta$ and $T_{e13}$, the greater the slopes. While in Fig.\ref{T1}(b), the larger $M_2$, the smaller the slope. In Fig.\ref{T1}, $M_2$, $M_{\tilde{L}12}^2$ and $M_{\tilde{E}}^2$ vary much less than their current limits in the $10^{-16}- 10^{-12}$ region. In summary, $M_2$,~$M_{\tilde{L}12}^2$, $M_{\tilde{E}12}^2$ and $M_{\tilde{E}}^2$ are sensitive parameters that have a significant effect on Br($\mu\rightarrow e\gamma\gamma$).

\begin{figure}[ht]
\setlength{\unitlength}{5mm}
\centering
\includegraphics[width=3in]{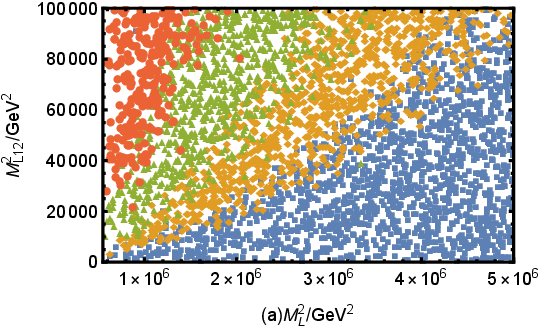}
\setlength{\unitlength}{5mm}
\centering
\includegraphics[width=2.9in]{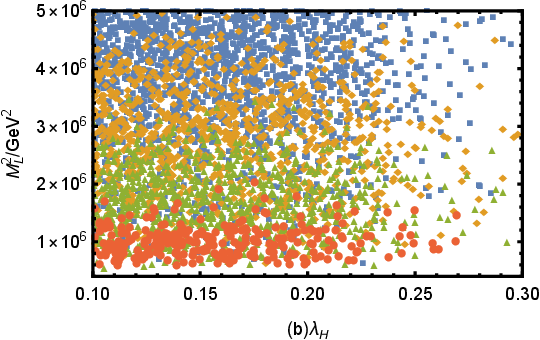}
\setlength{\unitlength}{5mm}
\centering\nonumber\\
\includegraphics[width=2.9in]{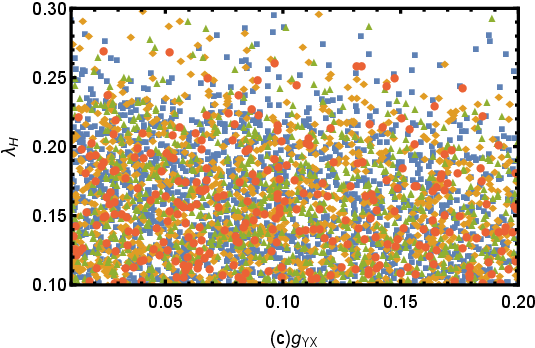}
\caption{Under the premise of current limit on LFV decay $\mu\rightarrow e\gamma\gamma$, reasonable parameter space is selected to scatter points, with the notation \textcolor{black-blue}{$\blacksquare$} $(0<Br(\mu\rightarrow e\gamma\gamma)<6\times 10^{-14})$, \textcolor{black-yellow}{$\blacklozenge$} $(6\times 10^{-14}\leq Br(\mu\rightarrow e\gamma\gamma<3\times 10^{-13})$, \textcolor{black-green}{$\blacktriangle$} $(3\times 10^{-13}\leq Br(\mu\rightarrow e\gamma\gamma<3\times 10^{-12})$, \textcolor{black-red}{$\bullet$}  $(3\times 10^{-12}\leq Br(\mu\rightarrow e\gamma\gamma<7.2\times 10^{-11})$. }{\label {T2}}
\end{figure}

\begin{table*}
\caption{Scanning parameters for Fig.\ref{T2}, Fig.\ref{T4} and Fig.\ref{T6}.}
\begin{tabular*}{\textwidth}{@{\extracolsep{\fill}}|l|l|l|l|l|l|l|l|l|l|@{}}
\hline
Parameters&$\tan\beta$&$g_X$&$g_{YX}$&$~~~\lambda_H$&$~~~\lambda_C$&$\mu/{\rm GeV}$&$M_2/~{\rm GeV}$ &$M_{\tilde{L}}^2/~{\rm GeV}^2$&$M_{\tilde{\nu}}^2/~{\rm GeV}^2$\\
\hline
Min&~~5&0.3&~0.01~&~~~~~0.1~~~&~~~~-0.3~~&~~~1000~~&~~~~700&~~~~4$\times10^{5}~~$& ~~3$\times10^{5}$~~\\
\hline
Max&~~50&0.6&~0.2~&~~~~~0.3~~~&~~~~-0.1~~&~~~1300~~&~~~~2500&~~~~5$\times10^{6}~~$& ~~5$\times10^{6}$~~\\
\hline
\end{tabular*}
\label{B1}
\end{table*}

\begin{table*}
\caption{Scanning parameters for Fig.\ref{T2}.}
\begin{tabular*}{\textwidth}{@{\extracolsep{\fill}}|l|l|l|l|@{}}
\hline
Parameters&$M_{\tilde{L}12}^2/~{\rm GeV}^2$~~~~~~~&$T_{e12}/~{\rm GeV}$~~~~~~~&$T_{\tilde{\nu}12}/~{\rm GeV}$~~~~~~~\\
\hline
Min&~~~0~~~~~~~~&~~~- 400 ~~~~~~~~&~~~- 400~~~~~~~~ \\
\hline
Max&~~$10^{5}~~~~~~~~$ & ~~~400~~~~~~~~ & ~~~400 ~~~~~~~~\\
\hline
\end{tabular*}
\label{B2}
\end{table*}

Next, we randomly scan some parameters, which we represent in a tabular form. Fig.\ref{T2} is obtained from the parameters shown in Table \ref {B1} and Table \ref {B2}. We use $\textcolor{black-blue}{\blacksquare}~(0<Br(\mu\rightarrow e\gamma\gamma)<6\times 10^{-14}),~\textcolor{black-yellow}{\blacklozenge}~(6\times 10^{-14}\leq Br(\mu\rightarrow e\gamma\gamma<3\times 10^{-13}),~ \textcolor{black-green}{\blacktriangle}~(3\times 10^{-13}\leq Br(\mu\rightarrow e\gamma\gamma<3\times 10^{-12}),~\textcolor{black-red}{\bullet}~3\times 10^{-12}\leq Br(\mu\rightarrow e\gamma\gamma<7.2\times 10^{-11})$ to represent the results for different parameter spaces for the $\mu\rightarrow e\gamma\gamma$ process respectively.

Analysis of the relationship between $M_{\tilde{L}12}^2$ and $M_{\tilde{L}}^2$ is shown in Fig.\ref{T2}(a). The overall trend is obvious. With \textcolor{black-red}{$\bullet$} mainly concentrated in the upper left corner, the outer layer is \textcolor{black-green}{$\blacktriangle$}, followed by \textcolor{black-yellow}{$\blacklozenge$}, and the rightmost is \textcolor{black-blue}{$\blacksquare$}. When $M_{\tilde{L}12}^2$ approaches $1\times10^5~{\rm GeV}^2$ and $M_{\tilde{L}}^2$ approaches $4\times10^5~{\rm GeV}^2$, Br($\mu\rightarrow e\gamma\gamma$) gets the maximum value. In Fig.\ref{T2}(b) we analyze the relationship between $M_{\tilde{L}}^2$ and $\lambda_H$, and we find that the trend of dispersion is weak, where \textcolor{black-red}{$\bullet$} part is mainly in $4\times 10^{5}<M_{\tilde{L}}^2\leq 1.4\times10^6$, \textcolor{black-green}{$\blacktriangle$} is mainly in $1.4\times 10^6<M_{\tilde{L}}^2\leq 2.4\times10^6$, \textcolor{black-yellow}{$\blacklozenge$} is mainly in $2.4\times 10^6<M_{\tilde{L}}^2\leq 4.2\times10^6$, \textcolor{black-blue}{$\blacksquare$} is mainly in $4.2\times 10^6<M_{\tilde{L}}^2\leq 5\times10^6$. The four color levels are obvious, and the value of Br($\mu\rightarrow e\gamma\gamma$) increases as $M_{\tilde{L}}^2$ decreases. Fig.\ref{T2}(c) shows the effects of $\lambda_H$ and $g_{YX}$ on Br($\mu\rightarrow e\gamma\gamma$). All points are mainly concentrated near the x-axis. We find that it is denser in the range of $0.1<\lambda_H\leq 0.2$ and increasingly sparse in the range of $0.2<\lambda_H\leq 0.3$.

\subsection{$\tau\rightarrow \mu\gamma\gamma$}
\begin{figure}[ht]
\setlength{\unitlength}{5mm}
\centering
\includegraphics[width=2.75in]{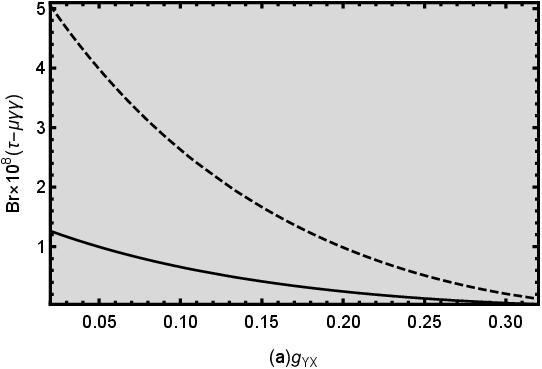}
\setlength{\unitlength}{5mm}
\centering
\includegraphics[width=2.9in]{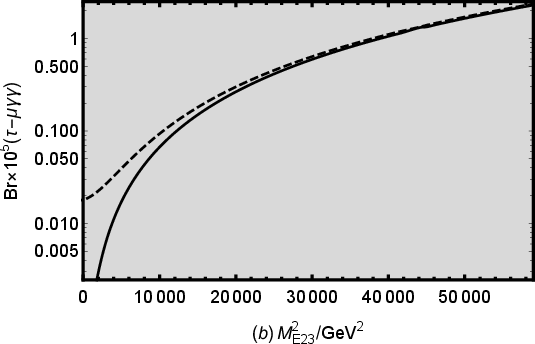}
\setlength{\unitlength}{5mm}
\centering\nonumber\\
\includegraphics[width=2.85in]{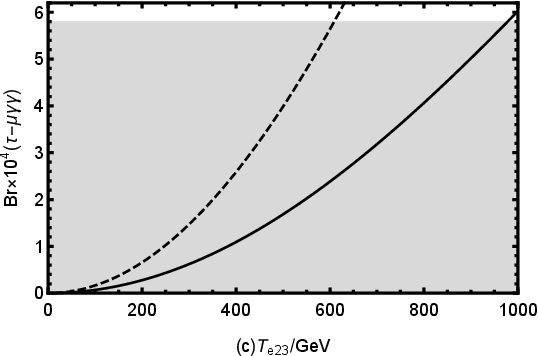}
\setlength{\unitlength}{5mm}
\centering
\includegraphics[width=2.9in]{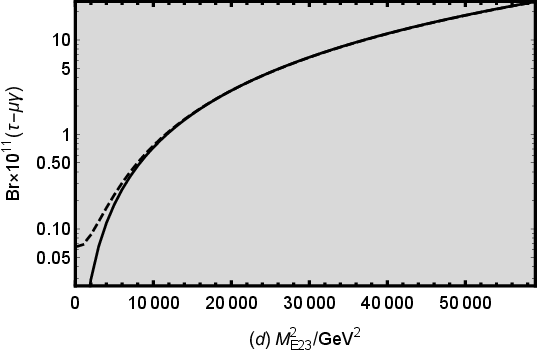}
\caption{Br($\tau\rightarrow \mu\gamma\gamma$) and Br($\tau\rightarrow \mu\gamma$)  diagrams affected by different parameters.
 The gray areas are reasonable value range. With $M_{\tilde{\nu}12}^2=100~{\rm GeV}^2$, the dashed and solid lines in Fig.\ref{T3}(a) correspond to $M_{\tilde{E}23}^2=2000~{\rm GeV}^2$ and $M_{\tilde{E}23}^2=1000~{\rm GeV}^2$. The dashed and solid lines in Fig.\ref{T3}(b)(d) correspond to $M_{\tilde{L}23}^2=2000~{\rm GeV}^2$ and $M_{\tilde{L}23}^2=200~{\rm GeV}^2$. As $M_{\tilde{L}23}^2=500~{\rm GeV}^2$, the dashed and solid lines in Fig.\ref{T3}(c) correspond to $M_2=1600~{\rm GeV}$ and $M_2=1200~{\rm GeV}$.}{\label {T3}}
\end{figure}

With the parameters $M_{\tilde{E}}^2=8\times10^5~{\rm GeV}^2$, $M_{\tilde{L}}^2=1\times10^6~{\rm GeV}^2$, $M_2$ = 1200~{\rm GeV}, $\tan\beta$ = 20, $g_X$=0.3, we plot the schematic diagram of Br($\tau\rightarrow \mu\gamma\gamma$) affected by different parameters.  The relationship between Br($\tau\rightarrow \mu\gamma$) and the parameter $M_{\tilde{E}23}^2$ is plotted in Fig.\ref{T3}(d) utilizing the same parameters as in Fig.\ref{T3}(b).

We study the effects of parameters $g_{YX}$, $M_{\tilde{E}23}^2$ and $T_{e23}$ on Br($\tau\rightarrow \mu\gamma\gamma$) in Fig.\ref{T3}. In Fig.\ref{T3}(a) we set $M_{\tilde{\nu}12}^2=100~{\rm GeV}^2$ and plot the relationship between Br($\tau\rightarrow \mu\gamma\gamma$) and $g_{YX}$, where the dashed line corresponds to $M_{\tilde{E}23}^2=2000~{\rm GeV}^2$ and the solid line corresponds to $M_{\tilde{E}23}^2=1000~{\rm GeV}^2$. The dashed line is larger than the solid line. We can clearly see that these two lines decrease as $g_{YX}$ increases. The solid and dashed lines are located in the gray area. In Fig.\ref{T3}(b)(d), the relationship between Br($\tau\rightarrow \mu\gamma\gamma$), Br($\tau\rightarrow \mu\gamma$) and $M_{\tilde{E}23}^2$ are shown, and the results are plotted with the dashed lines ($M_{\tilde{L}23}^2=2000~{\rm GeV}^2$) and the solid lines ($M_{\tilde{L}23}^2=200~{\rm GeV}^2$), respectively. We find that lines show an increasing trend, and their values gradually coincide as $M_{\tilde{E}23}^2$ increases. Both are smaller than the experimental upper limit. 

In Fig.\ref{T3}(c), Br($\tau\rightarrow \mu\gamma\gamma$) varies with $T_{e23}$ as $M_{\tilde{L}23}^2=500~{\rm GeV}^2$, the dashed and solid lines correspond to $M_2=1600~{\rm GeV}$ and $M_2=1200~{\rm GeV}$, respectively. It can be clearly seen that both the solid and dashed lines have a tendency to rise. The rising range of the dashed line is larger than that of the solid line. The dashed line with $T_{e23}$ range of $0 -600~{\rm GeV}$ and the solid line with $T_{e23}$ range of $0 -960~{\rm GeV}$ lie within the experimental limits. In summary the parameters $g_{YX}$, $M_{\tilde{E}23}^2$ and $T_{e23}$ have influence on Br($\tau\rightarrow \mu\gamma\gamma$) and are sensitive parameters.

\begin{figure}[ht]
\setlength{\unitlength}{5mm}
\centering
\includegraphics[width=3in]{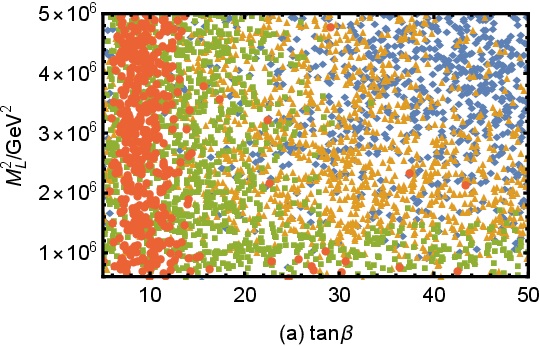}
\setlength{\unitlength}{5mm}
\centering
\includegraphics[width=2.9in]{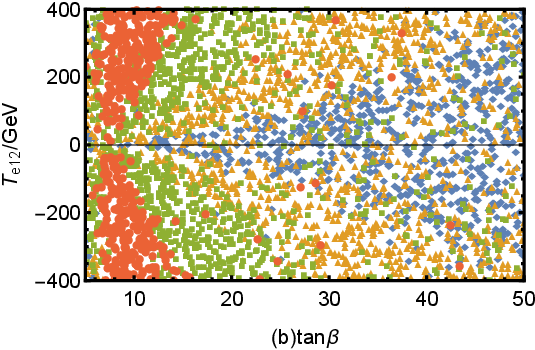}
\caption{Under the premise of current limit on LFV decay $\tau\rightarrow \mu\gamma\gamma$, reasonable parameter space is selected to scatter points, with the notation \textcolor{black-blue}{$\blacklozenge$} $(0<Br(\tau\rightarrow \mu\gamma\gamma)<4\times 10^{-6})$,~\textcolor{black-yellow}{$\blacktriangle$} $(4\times 10^{-6}\leq Br(\tau\rightarrow \mu\gamma\gamma)<4\times 10^{-5})$,~\textcolor{black-green}{$\blacksquare$} $(4\times 10^{-5}\leq Br(\tau\rightarrow \mu\gamma\gamma)<5.8\times 10^{-4})$,~\textcolor{black-red}{$\bullet$} $(Br(\tau\rightarrow \mu\gamma\gamma)\geq 5.8\times 10^{-4})$.}{\label {T4}}
\end{figure}

\begin{table*}
\caption{Scanning parameters for Fig.\ref{T4}}
\begin{tabular*}{\textwidth}{@{\extracolsep{\fill}}|l|l|l|l|@{}}
\hline
Parameters&$M_{\tilde{L}23}^2/~{\rm GeV}^2$~~~~~~~&$T_{e23}/~{\rm GeV}$~~~~~~~&$T_{\tilde{\nu}23}/~{\rm GeV}$~~~~~~~\\
\hline
Min&~~~0~~~~~~~~&~~~- 400 ~~~~~~~~&~~~- 400~~~~~~~~ \\
\hline
Max&~~$10^{5}~~~~~~~~$ & ~~~400~~~~~~~~ & ~~~400 ~~~~~~~~\\
\hline
\end{tabular*}
\label{B3}
\end{table*}

Next, we randomly scan some parameters, namely those shown in Table \ref {B1} and Table \ref {B3}, according to which we obtain Fig.\ref{T4}. We use \textcolor{black-blue}{$\blacklozenge$} $(0<Br(\tau\rightarrow \mu\gamma\gamma)<4\times 10^{-6})$,~\textcolor{black-yellow}{$\blacktriangle$} $(4\times 10^{-6}\leq Br(\tau\rightarrow \mu\gamma\gamma)<4\times 10^{-5})$,~\textcolor{black-green}{$\blacksquare$} $(4\times 10^{-5}\leq Br(\tau\rightarrow \mu\gamma\gamma)<5.8\times 10^{-4})$,~\textcolor{black-red}{$\bullet$} $(Br(\tau\rightarrow \mu\gamma\gamma)\geq 5.8\times 10^{-4})$ to denote the results in different parameter spaces in the process $\tau\rightarrow \mu\gamma\gamma$.

Analysis of the relationship between $\tan\beta$ and $M_{\tilde{L}12}^2$ is shown in Fig.\ref{T4}(a). We are able to clearly find that \textcolor{black-blue}{$\blacklozenge$} is mainly concentrated in the upper right corner. Then \textcolor{black-yellow}{$\blacktriangle$}, \textcolor{black-green}{$\blacksquare$} step by step, and \textcolor{black-red}{$\bullet$} is beyond the experimental limits ($Br(\tau\rightarrow \mu\gamma\gamma)\geq5.8\times10^{-4}$), and they are mainly concentrated in the range of $5<\tan\beta<14$. In Fig.\ref{T4}(b) we plot the relationship between $\tan\beta$ and $T_{e12}$ scattered points and find that each color is arranged like a triangle and is symmetric about $T_{e12}=0$. The \textcolor{black-blue}{$\blacklozenge$} is concentrated in $14<\tan\beta\leq50$. \textcolor{black-yellow}{$\blacktriangle$} is next to the blue color mainly in $7<\tan\beta\leq50$, followed by \textcolor{black-green}{$\blacksquare$} mainly in $8<\tan\beta\leq30$. The last \textcolor{black-red}{$\bullet$} beyond the experimental limit is mainly present in $6<\tan\beta\leq16$.

\subsection{$\tau\rightarrow e\gamma\gamma$ }
Using the parameters $M_{\tilde{E}}^2=8\times10^5~{\rm GeV}^2$, $M_{\tilde{\nu}ij}^2=0$ $(i,j=1,2,3,~~i\neq j)$, $g_{YX}=0.1$ we plot the diagrams of Br($\tau\rightarrow e\gamma\gamma$) affected by different parameters, and the same set of parameters is used to explore Br($\tau\rightarrow e\gamma$) and find the connection between them.

\begin{figure}[ht]
\setlength{\unitlength}{5mm}
\centering
\includegraphics[width=2.9in]{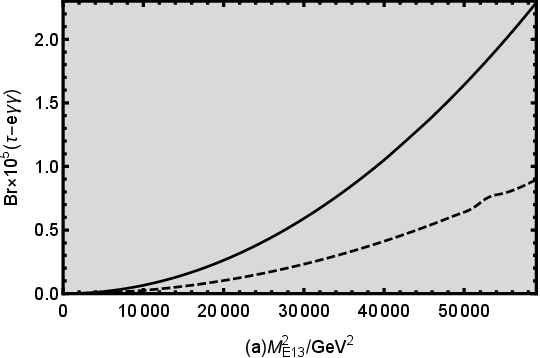}
\setlength{\unitlength}{5mm}
\centering
\includegraphics[width=2.9in]{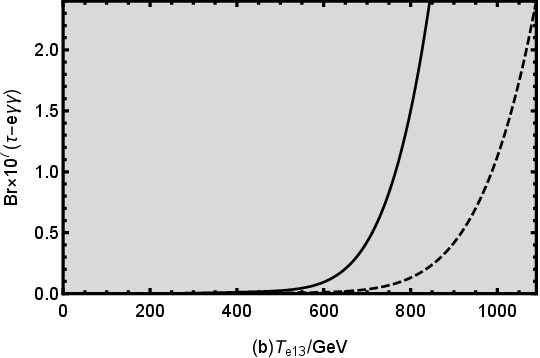}
\setlength{\unitlength}{5mm}
\centering
\includegraphics[width=2.95in]{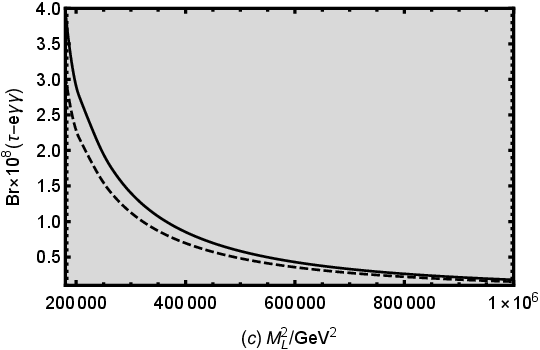}
\setlength{\unitlength}{5mm}
\centering
\includegraphics[width=2.85in]{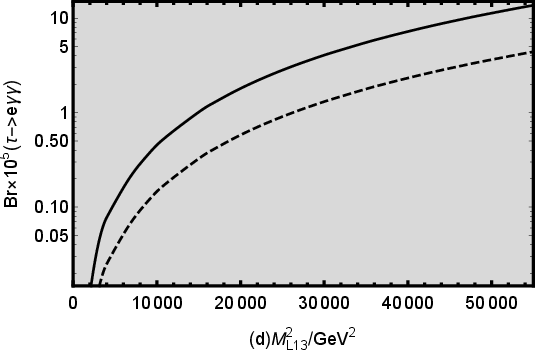}
\setlength{\unitlength}{5mm}
\centering
\includegraphics[width=2.9in]{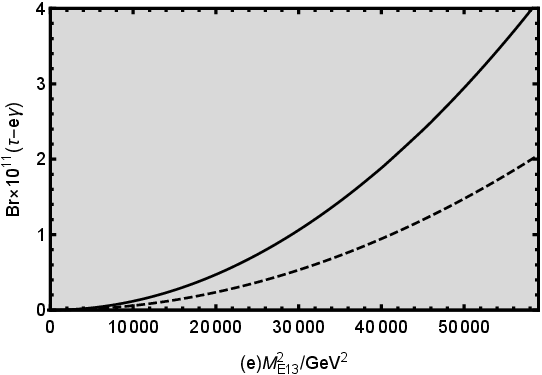}
\setlength{\unitlength}{5mm}
\centering
\includegraphics[width=2.9in]{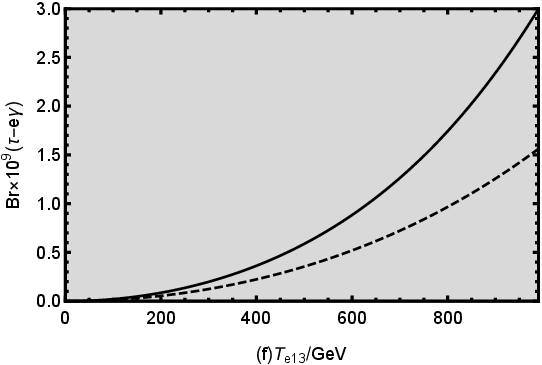}
\caption{{Br($\tau\rightarrow e\gamma\gamma$) and Br($\tau\rightarrow e\gamma$) schematic diagrams affected by different parameters. The gray areas are reasonable value range, where Br($\tau\rightarrow e\gamma\gamma$) and Br($\tau\rightarrow e\gamma$) are lower than the upper limit. The dashed and solid lines in Fig.\ref{T5}(a)(e) correspond to $g_X=0.4$ and $g_X=0.3$. The dashed and solid lines in Fig.\ref{T5}(b)(f) $\tan\beta=25$ and $\tan\beta=20$. The dashed and solid lines in Fig.\ref{T5}(c) correspond to $M_1=1400~{\rm GeV}$ and $M_1=1000~{\rm GeV}$}. In Fig.\ref{T5}(d), with $\tan\beta=15$, the dashed line corresponds to $\mu=1300~{\rm GeV}$ and the solid line corresponds to $\mu=1000~{\rm GeV}$.}{\label {T5}}
\end{figure}

We study the effects of the parameters $M_{\tilde{E}13}^2$,~$T_{e13}$, $M_{\tilde{L}13}^2$, $g_X$, $\tan\beta$, $M_1$, $\mu$ and $M_{\tilde{L}}^2$ on Br($\tau\rightarrow e\gamma\gamma$) in Fig.\ref{T5}. In Fig.\ref{T5}(a)(e), we plot the relationship between Br($\tau\rightarrow e\gamma\gamma$), Br($\tau\rightarrow e\gamma$) and $M_{\tilde{E}13}^2$, where the dashed lines correspond to $g_{X}=0.4$ and the solid lines correspond to $g_{X}=0.3$. The solid curve is larger than the dashed curve, and two lines increase with the increasing $M_{\tilde{E}13}^2$. The solid and dashed lines are located in the gray area. In Fig.\ref{T5}(b)(f), the relationship between Br($\tau\rightarrow e\gamma\gamma$), Br($\tau\rightarrow e\gamma$) and $T_{e13}$ are shown, and the results are plotted with the dashed lines ($\tan\beta=25$) and the solid lines ($\tan\beta=20$), respectively. In Fig.\ref{T5}(b), we find very small values for the solid line at $0<T_{e13}<500$ GeV and for the dashed line at $0<T_{e13}<600$ GeV, but then increase substantially. In Fig.\ref{T5}(c), let $M_{\tilde{L}ij}^2=200~{\rm GeV}^2$ ($i,j=1,2,3,~i\neq j$), with $M_1=1400~{\rm GeV}$ and $M_1=1000~{\rm GeV}$ corresponding to the dashed and solid lines, respectively, Br($\tau\rightarrow e\gamma\gamma$) varies with $M_{\tilde{L}}^2$. It can be clearly seen that both the solid and dashed lines show a decreasing trend. Let $g_X=0.3$, $\tan\beta=20$, $M_1=1000~{\rm GeV}$, in Fig.\ref{T5}(d), we plot Br($\tau\rightarrow e\gamma\gamma$) versus $M_{\tilde{L}13}^2$. The dashed and solid lines indicate $\mu=1300~{\rm GeV}$, $\mu=1000~{\rm GeV}$, respectively. It is clear that Br($\tau\rightarrow e\gamma\gamma$) becomes larger as $M_{\tilde{L}13}^2$ increases and smaller as $\mu$ increases.
In summary the parameters $M_{\tilde{E}13}^2$, $T_{e13}$, $M_{\tilde{L}13}^2$, $\mu$, $M_1$, $g_X$, $\tan\beta$ and $M_{\tilde{L}}^2$ have some influence on Br($\tau\rightarrow e\gamma\gamma$) and are the sensitive parameters.

\begin{figure}[ht]
\setlength{\unitlength}{5mm}
\centering
\includegraphics[width=2.9in]{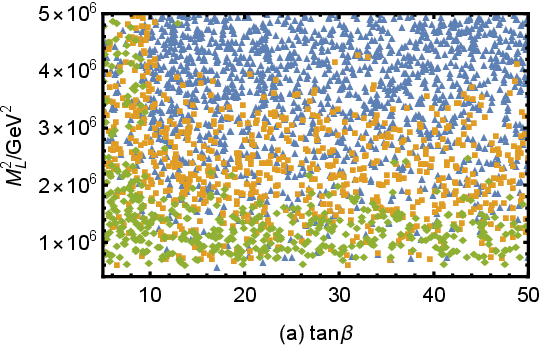}
\setlength{\unitlength}{5mm}
\centering
\includegraphics[width=2.9in]{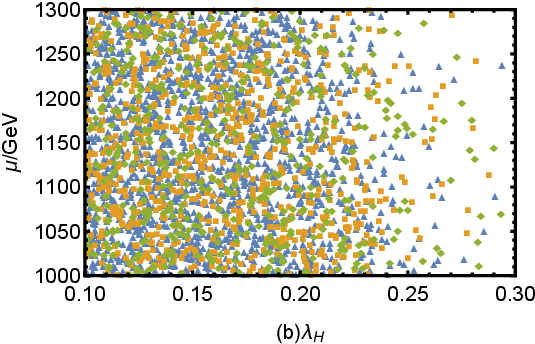}
\setlength{\unitlength}{5mm}
\centering\nonumber\\
\includegraphics[width=2.9in]{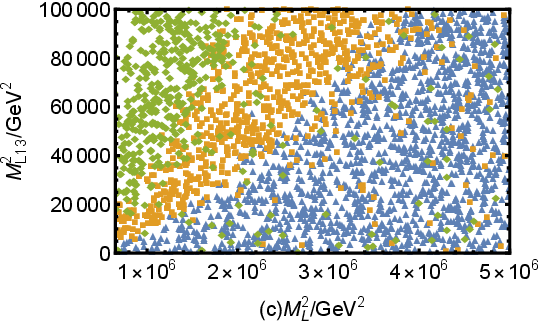}
\caption{Under the premise of current limit on LFV decay $\tau\rightarrow e\gamma\gamma$, reasonable parameter space is selected to scatter points, with the notation \textcolor{black-blue}{$\blacktriangle$} $(0<Br(\tau\rightarrow e\gamma\gamma)<2\times 10^{-6})$, \textcolor{black-yellow}{$\blacksquare$} $(2\times 10^{-6}\leq Br(\tau\rightarrow e\gamma\gamma)<2\times 10^{-5})$, \textcolor{black-green}{$\blacklozenge$} $(2\times 10^{-5}\leq Br(\tau\rightarrow e\gamma\gamma)<2.5\times 10^{-4})$.}{\label {T6}}
\end{figure}

\begin{table*}
\caption{Scanning parameters for Fig.\ref{T6}}
\begin{tabular*}{\textwidth}{@{\extracolsep{\fill}}|l|l|l|l|@{}}
\hline
Parameters&$M_{\tilde{L}13}^2/~{\rm GeV}^2$~~~~~~~&$T_{e13}/~{\rm GeV}$~~~~~~~&$T_{\tilde{\nu}13}/~{\rm GeV}$~~~~~~~\\
\hline
Min&~~~0~~~~~~~~&~~~- 400 ~~~~~~~~&~~~- 400~~~~~~~~ \\
\hline
Max&~~$10^{5}~~~~~~~~$ & ~~~400~~~~~~~~ & ~~~400 ~~~~~~~~\\
\hline
\end{tabular*}
\label{B4}
\end{table*}
Next, we randomly scan some parameters, namely those shown in Table \ref {B1} and Table \ref {B4}, according to which we obtain Fig.\ref{T6}. We use \textcolor{black-blue}{$\blacktriangle$} $(0<Br(\tau\rightarrow e\gamma\gamma)<2\times 10^{-6})$, \textcolor{black-yellow}{$\blacksquare$} $(2\times 10^{-6}\leq Br(\tau\rightarrow e\gamma\gamma)<2\times 10^{-5})$, \textcolor{black-green}{$\blacklozenge$} $(2\times 10^{-5}\leq Br(\tau\rightarrow e\gamma\gamma)<2.5\times 10^{-4})$ to denote the results in different parameter spaces in the process $\tau\rightarrow e\gamma\gamma$.

In Fig.\ref{T6}(a) we analyze the relationship between $\tan\beta$ and $M_{\tilde{L}}^2$. We are able to clearly find \textcolor{black-blue}{$\blacktriangle$} is mainly concentrated in the upper right position, \textcolor{black-yellow}{$\blacksquare$} is mainly concentrated in the middle position, and \textcolor{black-green}{$\blacklozenge$} is mainly concentrated in the bottom. The value of the branching ratio shows an increasing trend from the upper right to the lower left. In Fig.\ref{T6}(b) we analyze the relationship between $\lambda_H$ and $\mu$. The points are very dense in the range of $0.1<\lambda_H\leq0.2$ and start to become sparse after $\lambda_H>0.2$. In Fig.\ref{T6}(c) we analyze the relationship between $M_{\tilde{L}}^2$ and $M_{\tilde{L}13}^2$. 
We find that the \textcolor{black-green}{$\blacklozenge$} is mainly concentrated in $4\times10^5~{\rm GeV}^2 <M_{\tilde{L}}^2<1.8\times10^6~{\rm GeV}^2$(view from the above axis)and $2\times10^4~{\rm GeV}^2<M_{\tilde{L}13}^2<1\times10^5~{\rm GeV}^2$(view from the left axis), \textcolor{black-yellow}{$\blacksquare$} is mainly concentrated in $5000~{\rm GeV}^2<M_{\tilde{L}}^2<3.8\times10^6~{\rm GeV}^2$(view from the above axis) and $0 <M_{\tilde{L}13}^2<1\times10^5~{\rm GeV}^2$(view from the left axis), \textcolor{black-blue}{$\blacktriangle$} is mainly concentrated in
$4\times10^5~{\rm GeV}^2 <M_{\tilde{L}}^2<3.8\times10^6~{\rm GeV}^2$(view from the above axis) and $0 <M_{\tilde{L}13}^2<1\times10^5~{\rm GeV}^2$.

\section{Conclusion}
From the order of magnitude of branching ratios and data analysis, we know that the $\mu\rightarrow e\gamma$  process is more restrictive to LFV, based on which we investigate the LFV process of $l_j\rightarrow l_i \gamma\gamma$ in this paper. We have made use of the $U(1)_X$SSM, which has contribute to our study.
We consider the Feynman diagrams of $l_j\rightarrow l_i \gamma\gamma$ and
perform extensive calculations to draw line diagrams of the different
parameters versus Br($l_j\rightarrow l_i \gamma\gamma$), followed
by a large scan of the parameters. Numerical results show that Br($\mu\rightarrow e\gamma\gamma$),
Br($\tau\rightarrow \mu\gamma\gamma$) and Br($\tau\rightarrow e\gamma\gamma$) are related
 to the leptonic flavor mixing parameters. By analyzing the values, we learn that the branching ratios can reach $10^{-12}$ for $\mu\rightarrow e\gamma\gamma$, $10^{-4}$ for $\tau\rightarrow \mu\gamma\gamma$, and $10^{-5}$ for $\tau\rightarrow e\gamma\gamma$. The branching ratios of the three processes in $U(1)_X$SSM are very close to or even exceed their respective experimental upper limits. This provides a reference for other future work on leptonic flavor destruction.

We consider the constraints on the LFV branching ratio for $l_j\rightarrow l_i \gamma\gamma$. In the numerical calculation, we include many parameters as variables, including $\tan\beta$, $g_X$, $g_{YX}$, $\lambda_H$, $\lambda_C$, $\mu$, $M_2$, $M_{\tilde{L}}^2$, $M_{\tilde{\nu}}^2$, $M_{\tilde{L}ij}^2$, $T_{\nu ij}$ and $T_{eij}$. By analyzing the numerical results, we find that a part of parameters have a great influence on the results. $M_{\tilde{L}}^2$, $M_{\tilde{E}}^2$, $g_{X}$, $g_{YX}$, $M_2$ and $\tan\beta$ are important parameters. $M_{\tilde{L}ij}^2$, $T_{eij}$, $M_{\tilde{E}ij}^2$, $M_{\tilde{\nu}ij}^2$ and $T_{\nu ij}$ are sensitive parameters. We also make comparison with certain parameters of the $l_j\rightarrow l_i\gamma$ processes and obtain that our work can satisfy the restrictions, which is very meaningful. To save space in the paper, we only compare and explain some parameters of the $l_j\rightarrow l_i\gamma$ processes. However, through comprehensive research, we prove that all parameters in our work satisfy the limitations of the $l_j\rightarrow l_i\gamma$ processes. In summary, we find that many parameters have a greater or lesser effect on LFV, but the non-diagonal elements corresponding to the initial and final leptons generation are the most sensitive to LFV. So we believe that the main sensitive parameters and sources leading to LFV are the
 non-diagonal elements involving the initial and final leptons.
 This work can benefit the detection of new physics.

\begin{acknowledgments}

This work is supported by National Natural Science Foundation of China (NNSFC)(No.12075074),
Natural Science Foundation of Hebei Province(A2020201002, A2023201040, A202201022, A2022201017, A2023201041),
Natural Science Foundation of Hebei Education Department (QN2022173),
Post-graduate's Innovation Fund Project of Hebei University (HBU2023SS043, HBU2024SS042), the youth top-notch talent support program of the Hebei Province.
\end{acknowledgments}

\end{document}